\documentclass[preprints,article,accept,moreauthors,pdftex]{Definitions/mdpi} 

\renewcommand{\contentleftcolumn}{}

\firstpage{1} 
\pubvolume{1}
\issuenum{1}
\articlenumber{0}
\pubyear{ }
\copyrightyear{ }
\datereceived{ } 
\daterevised{ } 
\dateaccepted{ } 
\datepublished{ } 
\hreflink{https://doi.org/} 

\Title{Temperature dependence of nonlinear elastic moduli of polystyrene}

\TitleCitation{Temperature dependence of nonlinear elastic moduli of polystyrene}

\Author{ Andrey V. Belashov, Anna A. Zhikhoreva,   Yaroslav M. Beltukov and Irina V. Semenova$^{*}$}

\AuthorNames{ A.V. Belashov, A.A. Zhikhoreva,  Y.M. Beltukov and I.V. Semenova}

\AuthorCitation{Belashov, A.V.; Zhikhoreva, A.A.;  Beltukov, Y.M., Semenova, I.V.}

\address{%
 \quad Ioffe Institute, Russian Academy of Sciences, 26, Polytekhnicheskaya, St. Petersburg, 194021, Russia
}

\corres{Correspondence: irina.semenova@mail.ioffe.ru}

\abstract{Nonlinear elastic properties of polymers and polymeric composites are essential for accurate prediction of their response to dynamic loads, which is crucial in a wide range of applications. These properties can be affected by strain rate, temperature, and pressure. The temperature susceptibility of nonlinear elastic moduli of polymers remains poorly understood. We have recently observed a significant frequency dependence of the nonlinear elastic (Murnaghan) moduli of polystyrene. In this paper we expand this analysis by the temperature dependence. The measurement methodology was based on the acousto-elastic effect, and involved analysis of the dependencies of velocities of longitudinal and shear single-frequency ultrasonic waves in the sample on the applied static pressure. Measurements were performed at different temperatures in the range of 25--65 $^{\circ}$C and at different frequencies in the range of 0.75--3 MHz. The temperature susceptibility of the nonlinear moduli $l$ and $m$ was found to be two orders of magnitude larger than that of linear moduli $\lambda$ and $\mu$. At the same time, the observed variations of $n$ modulus with temperature were low and within the measurement tolerance. The observed tendencies can be explained by different influence of pressure on relaxation processes in the material at different temperatures.} 


\begin{document}



\section{Introduction}

Polymers are a class of materials characterized by long chains of repeating molecular units, which impart unique mechanical, thermal, and chemical properties. Their versatility and adaptability make them essential in a wide range of applications, from everyday consumer products to advanced industrial components~\cite{Grosberg-giant-molecules-here-2011, Osswald-material-science-polymers-2012}. The ability to tailor the properties of polymers through chemical modifications and processing techniques has led to their widespread use in various fields, including packaging, automotive, aerospace, electronics, and healthcare.

In recent years, the incorporation of nanomaterials into polymer matrices has given rise to polymer nanocomposites, which exhibit significantly enhanced properties compared to their conventional polymer counterparts~\cite{Mai-polymer-nanocomposites-2006, Peters-handbook-composites-1998}. By integrating nanoscale fillers, such as nanoparticles, nanotubes, or nanofibers, into the polymer matrix, one can achieve improvements in mechanical strength, thermal stability, barrier properties, and electrical conductivity. These enhancements are primarily attributed to the high surface area-to-volume ratio of nanomaterials, which facilitates better interaction with the polymer matrix and leads to a more effective load transfer at the nanoscale. Polymer nanocomposites enable the development of lightweight materials that can meet the demanding performance requirements of modern applications. For instance, in the automotive and aerospace industries, reducing weight while maintaining structural integrity is crucial for improving fuel efficiency and overall performance. Additionally, the unique properties of polymer nanocomposites make them suitable for innovative applications in fields such as energy storage, environmental remediation, and biomedical devices.

Despite the numerous advantages offered by polymers and polymer nanocomposites, understanding their mechanical behavior under various loading conditions remains a critical area of research~\cite{Osswald-material-science-polymers-2012, Siviour-high-strain-rate-2016}. The mechanical properties of polymer materials are significantly affected by strain rate, temperature, and pressure. Much of the existing research was focused primarily on the \emph{linear} elastic characteristics of these materials. The frequency and temperature dependence of the linear elastic moduli have been studied in various polymers and polymer nanocomposites~\cite{
Capodagli-isothermal-viscoelastic-properties-2008,
Yadav-effect-thermomechanical-couplings-2020,
Ionita-prediction-polyurethane-behaviour-2020,
Arrigo-rheological-behavior-polymer-2020,
Filippone-unifying-approach-linear-2012,
Hu-temperature-frequency-dependent-2013,
Wang-linear-nonlinear-viscoelasticity-2016}.

The study of \emph{nonlinear} elastic properties has become increasingly important in recent years, as it provides critical insights into the behavior of materials under various loading conditions. Nonlinear elasticity refers to the phenomenon where the stress-strain relationship deviates from linearity, particularly at higher stress levels~\cite{Fu-nonlinear-elasticity-theory-2001}. This behavior is essential for an accurate prediction of the material response to dynamic loads, which is crucial in applications ranging from structural engineering to geophysics and biophysics~\cite{
Astorga-nonlinear-elasticity-observed-2018,
Beresnev-nonlinear-soil-response-1996,
Field-nonlinear-groundmotion-amplification-1997,
Renaud-situ-characterization-shallow-2014,
Ogden-nonlinear-elasticity-anisotropy-2003,
Storm-nonlinear-elasticity-biological-2005}.
Understanding the nonlinear elastic properties is particularly important for polymers and other soft materials, since they may experience large deformations under relatively small applied forces, necessitating a departure from classical linear elasticity models~\cite{Philipp-immense-elastic-nonlinearities-2013}.

Several methods have been developed for  experimental evaluation of nonlinear elastic properties of materials. These techniques include dynamic acousto-elasticity~\cite{Hughes-secondorder-elastic-deformation-1953, Riviere-pump-probe-waves-2013, Lott-nonlinear-elasticity-rocks-2017}, second harmonic generation~\cite{Matlack-review-second-harmonic-2015}, Brillouin scattering~\cite{Kruger-nonlinear-elastic-properties-1991}, coda wave interferometry~\cite{Payan-determination-third-order-2009}, strain solitary waves~\cite{Garbuzov-determination-thirdorder-elastic-2016}, as well as Lamb~\cite{Bermes-experimental-characterization-material-2007} and Rayleigh~\cite{Masurkar-analyzing-features-material-2019} waves. Nonlinear elastic properties of materials can be described by a set of nonlinear elastic moduli, which characterize the deviation of a stress-strain relationship from linear behavior. Different sets have been proposed by Landau and Lifshitz~\cite{Landau}, Murnaghan~\cite{Murnaghan}, and Thurston and Brugger~\cite{Thurston-thirdorder-elastic-constants-1964}. The nonlinear elastic moduli defined in these models are interrelated and can be derived from one another. In particular, Murnaghan's model describes the nonlinear elastic behavior of isotropic solid materials using three third-order elastic moduli (denoted as $l$, $m$, and $n$) along with two second-order Lame moduli ($\lambda$ and $\mu$). These third-order elastic moduli, along with their linear combinations, are valuable for predicting fatigue damage~\cite{Nagy-fatigue-damage-assessment-1998}, understanding thermoelastic properties of crystalline solids~\cite{Wallace-thermoelastic-theory-stressed-1970}, analyzing radiation damage~\cite{Matlack-sensitivity-ultrasonic-nonlinearity-2014}, and studying phenomena such as creep and thermal aging~\cite{Matlack-review-second-harmonic-2015}. Overall, nonlinear parameters have been found to be more responsive to structural changes in materials compared to their linear counterparts~\cite{Jhang-nonlinear-ultrasonic-techniques-2009}.

The significant temperature dependence of the nonlinear elastic moduli has been observed in vitreous silica~\cite{Wang-temperature-dependences-thirdorder-1992}, metal-matrix composites~\cite{Mohrbacher-temperature-dependence-thirdorder-1993}, and constrained steel blocks~\cite{Nucera-nonlinear-wave-propagation-2014}. Nonlinear elastic moduli demonstrated significant changes at the phase transition temperature~\cite{Meeks-temperature-dependence-thirdorder-1970}. Exploitation of the thermal susceptibility of nonlinear attributes of elastic waves has ushered a new avenue to enhance the accuracy, precision, and reliability of the use of acoustic nonlinearity in elastic wave imaging~\cite{Wang-advancing-elastic-wave-2020}.
However, the temperature susceptibility of nonlinear elastic moduli in polymers remains poorly understood. 

Polystyrene, a commercially significant thermoplastic polymer, is known for its excellent mechanical properties, optical clarity, and chemical resistance. 
In prior investigations employing the acoustoelastic effect, we observed considerable variations in the nonlinear elastic moduli of polystyrene when different nanofillers were added to the polymer matrix~\cite{Belashov-relative-variations-nonlinear-2021}. Furthermore, our findings indicated a significant frequency dependence of the nonlinear elastic moduli of pure polystyrene~\cite{Belashov-frequency-dependence-nonlinear-2024}, which may be attributed to sub-MHz relaxation processes within the material. In this paper, we applied the developed methodology to analyze the dependence of nonlinear elastic moduli of polystyrene both on frequency in 0.7--3 MHz range and temperature in 25--65 $^{\circ}$C range.

\section{Material and Method}

Experiments were performed on block samples of commercial styrene copolymer containing 10\% of ethylene glycol dimethacrylate (EGDMA), produced by  Dzerzhinsk Enterprise for Organic Synthesis. Further in the paper we will refer to it as polystyrene (PS). The samples were fabricated in the form of 50 mm long bars, 10 x 10 mm$^{2}$ in cross section.

The methodology applied for measurements of nonlinear elastic moduli of the material was based on the acousto-elastic effect, the details are described in our previous papers \cite{Belashov-relative-variations-nonlinear-2021,Belashov-frequency-dependence-nonlinear-2024}.
Briefly, we analyzed the dependence of the velocities of longitudinal and shear single-frequency ultrasonic waves propagating in the sample on the applied static pressure at different temperatures of the sample. 
Application of the static pressure of the order of several MPa to the sample caused changes in the time of wave propagation in the sample. The obtained dependencies of velocities of longitudinal and shear waves on the applied pressure allowed for calculation of the set of nonlinear elastic moduli, as shown below.

For measurements of variations of the moduli values with temperature, we used the previously developed setup \cite{Belashov-frequency-dependence-nonlinear-2024} with some minor modifications. The setup comprised a jaw vice, a high-precision stress gauge,  a set of piezoelectric transducers used for generation and detection of longitudinal and shear ultrasonic waves, a pulse generator, an oscilloscope, and heating elements attached to the jaws.
The pressure applied to the sample was varied within the range of 0--16 MPa and was monitored  by the stress gauge. Continuous control was required, because temperature variations of the sample could also influence the applied pressure.

The temperature dependencies of nonlinear elastic moduli of polystyrene were analyzed in the temperature range of 25--65 $^{\circ}$C. This range guaranteed maintaining the sample in the glassy state even with regard to heating inhomogeneities. 
The glass transition temperature of polystyrene is about 100 $^{\circ}$C, however, softening occurs below this temperature. As shown by Lamberson et al. \cite{Lamberson1972}, reproducible results can be obtained in velocity measurements for this material at temperatures below 75 $^{\circ}$C. 

\begin{figure}[H]
\includegraphics[width=14 cm]{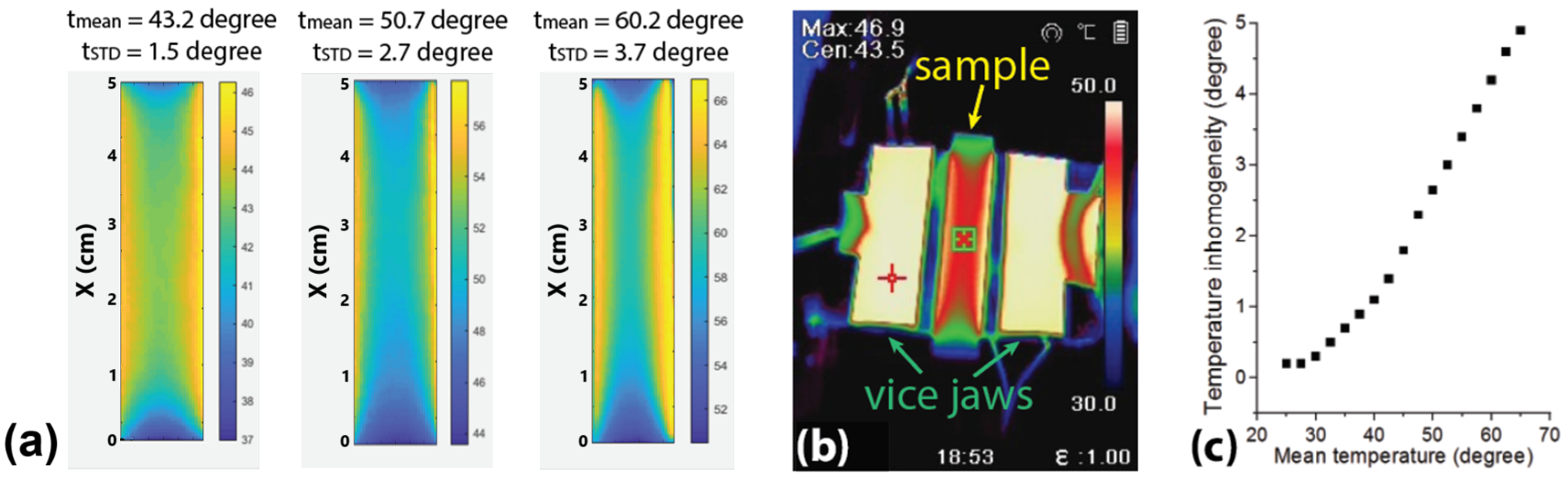}
\caption{(a) examples of spatial distributions of temperature on the sample surface at different mean temperatures (indicated on top of each image). (b) example of the thermal image of the sample and vise jaws, (c) dependence of temperature inhomogeneity in the sample on the mean temperature. \label{Setup}}
\end{figure}

Spatial distributions of  temperature in the sample were monitored using a thermal imager B20 (HIKMICRO, China). The imager was calibrated by comparing its readings with those from thermocouple sensors attached to the sample. So far as the piezoelectric transducers providing generation and detection of ultrasonic waves were applied to the input and output cross-sections of the sample, a completely uniform temperature distribution could not be obtained. Utilization of the thermal imager for analysis of the sample temperature allowed for both determination of the average temperature of the sample in the course of its heating, and assessment of temperature heterogeneity over the sample surface due to local heating from the jaws and contact with piezoelectric sensors. The observed temperature heterogeneity rose from 0.2 $^{\circ}$C at temperatures close to room to 5 $^{\circ}$C at high temperatures of about 65 $^{\circ}$C (Fig. \ref{Setup} (c)).
Figure \ref{Setup} (a) shows temperature maps of a polystyrene sample at different mean temperatures.

\begin{figure}[h!]
	\centering
	\includegraphics[width=13.5cm]{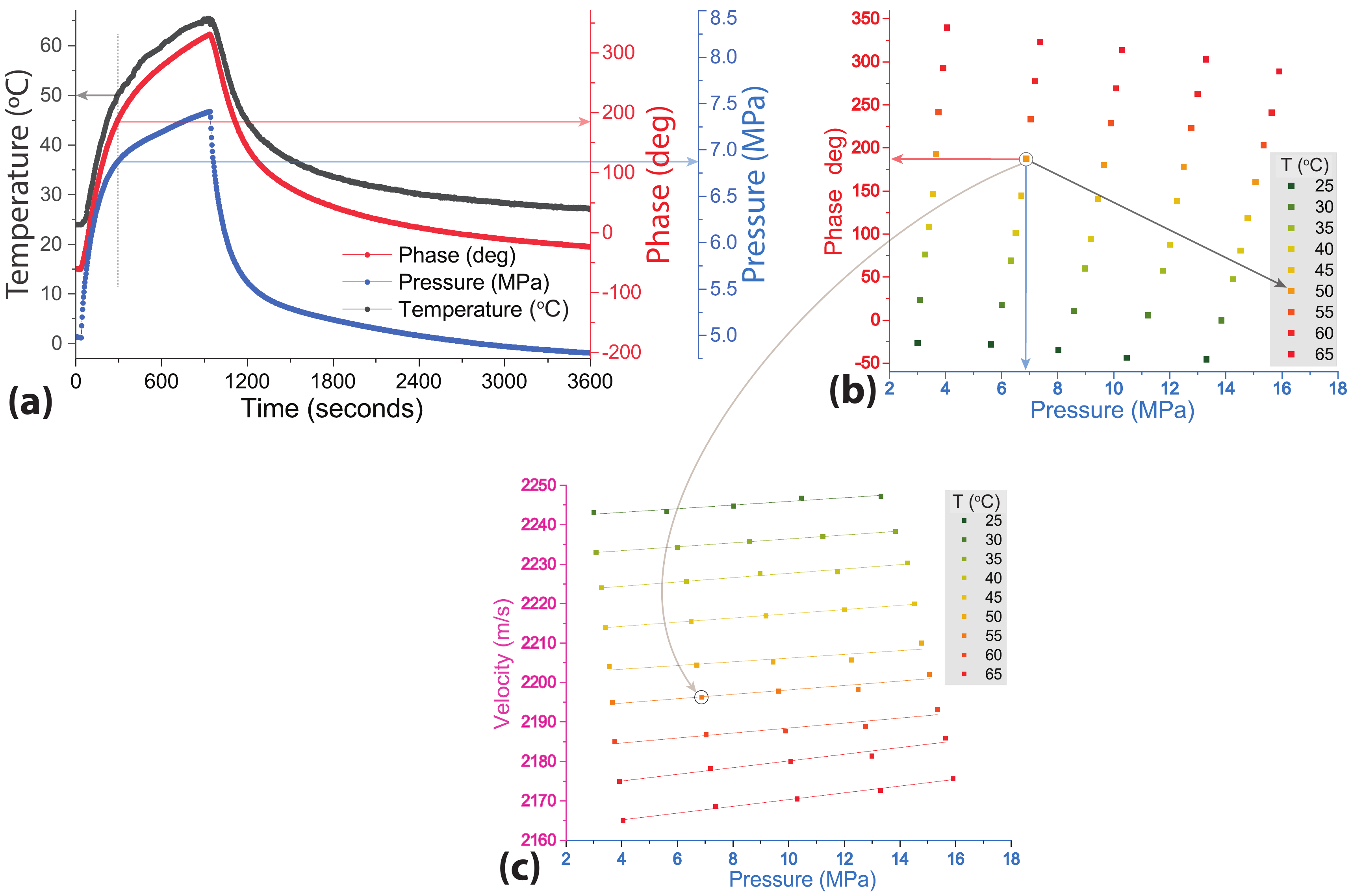}
	\caption{(a) changes in temperature, pressure and phase of the longitudinal ultrasonic wave at the frequency of 1.5 MHz and the initial pressure on the sample (pressure at room temperature) of 7.5 MPa. (b) phase of the detected ultrasonic wave at different pressures and temperatures of the sample, (c) dependence of the velocity of longitudinal ultrasonic waves at 1.5 MHz on the static pressure applied on the sample and sample temperature.}
	\label{PressureTempPhaseVariation}
\end{figure}

Experiments on the determination of nonlinear elastic moduli at different sample temperatures involved measurements of the velocities of three types of ultrasonic waves (longitudinal and shear with perpendicular directions) at varying pressure and temperature of the sample. Measurements of the phase shift $\Delta \phi$ of a sinusoidal signal at different temperatures $T$ and pressures $P$ at a specific frequency of ultrasonic waves  $f$ allowed us to determine the change in signal transit time as a function of temperature or pressure $\Delta t = \frac{\Delta \phi} {2 \pi f}$. Measurements could be carried out either by varying the pressure on the sample at several constant temperature values, or by changing the temperature of the sample at several constant static pressures. Approbation of these experimental scenarios has shown that quite a long time, up to 40--50 minutes, is required to stabilize the sample temperature (when the system enters a stationary temperature mode) and minimize its fluctuations during the measurement process. At the same time, reliable measurements require high temperature stability during pressure changes, which is why it was taking a significant amount of time to conduct experiments using this method. In addition, an increase in pressure on the sample due to its compression led to a tighter fit of the vice jaws and an increase in the heat transfer rate, which could cause the system leaving the stationary mode and temperature fluctuations. In this regard, in our experiments the sample was gradually heated from 25 to 65 $^{\circ}$C at each static pressure applied to the sample, $P \approx $ [3, 6, 9, 12, 15]  MPa. In the course of heating the sample temperature $T$, the phase of the recorded ultrasonic wave $\phi$ and the pressure were monitored. Note that the pressure rose slightly during heating due to sample expansion (see Fig. \ref{PressureTempPhaseVariation}(a)).

The data sets obtained at different static pressures and different temperatures $T \in $ [25, 30, \ldots, 60, 65] $^{\circ}$C provided values of the ultrasonic wave phase and its dependence on the applied pressure. These dependencies were constructed for each temperature: $\Delta \phi (P,T_1)$, $\Delta \phi (P,T_2)$ \ldots (see Fig. \ref{PressureTempPhaseVariation}(b)). The velocities of ultrasonic waves as a function of applied pressure for each specific temperature $T_i$ were calculated using the equation:
\begin{equation}\label{VelCalc_Ph}
    V(P, T_i) = \frac{L \cdot V(0,T_i)}{\Delta t (P, T_i) \cdot V(0,T_i) + L},
\end{equation}
where $V(0,T_i)$ is the velocity at zero  pressure, the change in the wave propagation time $\Delta t (P, T) = \frac{\Delta \phi (P, T)} {2 \pi f}$, and $L$ is the length of the part of the sample to which pressure was applied. The data obtained for longitudinal ultrasonic waves are plotted in Fig. \ref{PressureTempPhaseVariation}(c).

It is worth noting that the ultrasonic wave velocity in the sample also depends on the temperature. Therefore, it is necessary to apply in Eq.~(\ref{VelCalc_Ph}) the values of $V(0,T_i)$ for each specific temperature $T_i$. The dependence of velocity on pressure at a given temperature does not actually take into account the slight elongation of the sample with increasing pressure, the correct value of the velocity is $V_j = V_i(P)(1 + \varepsilon_x)$, where $\varepsilon_x = \nu P/E$ is the sample elongation in $x$ direction caused by the applied pressure. The sample elongation due to its compression is taken into account further when using Eqs.~(\ref{eq:l})--(\ref{eq:n}). The procedure used for measuring the dependence of ultrasonic wave velocity on pressure was also described in detail in our recent paper \cite{Belashov-frequency-dependence-nonlinear-2024}.

The obtained dependencies of the ultrasonic wave velocities on pressure allowed calculating the elastic moduli $M_j(P,T_i)= \rho \cdot (V_j(P,T_i))^2$  for three types of waves: $j=x$ for longitudinal waves, $j=y$ and $j=z$ for shear waves with the displacement vector parallel and perpendicular to the direction of applied pressure, respectively. 

Fitting of linear dependencies of the effective moduli $M_j(P,T_i)$ obtained at the temperature $T_i$ allowed us to determine the dimensionless slope coefficients $\alpha_j$ corresponding to each of the three types of waves. The nonlinear Murnaghan moduli $l,m,n$ were then calculated from the data obtained using the equations~\cite{Belashov-relative-variations-nonlinear-2021}:
\begin{gather}
    l = -\frac{3 \lambda +2 \mu }{2} \alpha_x-\frac{\lambda  (\lambda +\mu ) }{\mu}(1+2 \alpha_y)+\frac{\lambda^2 }{2 \mu}(1-2 \alpha_z),  \label{eq:l}\\
    m = -2 (\lambda +\mu ) \left(1+\alpha_y\right)+\lambda  \left(1-\alpha_z\right),  \label{eq:m}\\
    n = -4 \mu \left(1+\alpha_y-\alpha_z\right),  \label{eq:n}
\end{gather}
where $\lambda$ and $\mu$ are linear elastic Lame moduli, which were measured at zero pressure at each specific temperature, as described in~\cite{Belashov-relative-variations-nonlinear-2021}.
It is worth noting that at zero pressure the moduli $M_j(0,T_i)$ depend only on Lame moduli $\lambda$ and $\mu$:
\begin{align}
    M_x(0,T_i) &= \lambda +  2\mu, \label{eq:Mx}\\
    M_y(0,T_i) &= M_z(0,T_i) = \mu. \label{eq:My}
\end{align}

\section{Results and Discussion}

\subsection{Temperature dependence of ultrasonic wave velocities at zero pressure}

First, we performed an analysis of temperature and frequency dependencies of the velocity of longitudinal and shear ultrasonic waves in the sample at zero pressure. This data was necessary for further usage in calculating the linear ($\lambda, \mu$) and nonlinear ($l,m,n$) elastic moduli using Eqs.~(\ref{eq:l})--(\ref{eq:n}). 
In experiments, sinusoidal signals with a selected carrier frequency modulated by a Gaussian profile of finite width were generated, and the time of their propagation through the sample was measured at different temperatures in the range of 25--65 $^{\circ}$C.
Note that too small width (in terms of time) of the Gaussian function led to an increase in the spectral width of the wave packet and its distortion in the course of propagation in the sample; the too large width did not allow the boundaries of the wave packet to be clearly defined due to smooth increase and decrease in the amplitude of the modulated sinusoidal signal. 
Therefore, special care was taken for the selection of optimal parameters of a generated wave packet for each wave type.

\begin{figure}[H]
\centering
     \includegraphics[width=14cm]{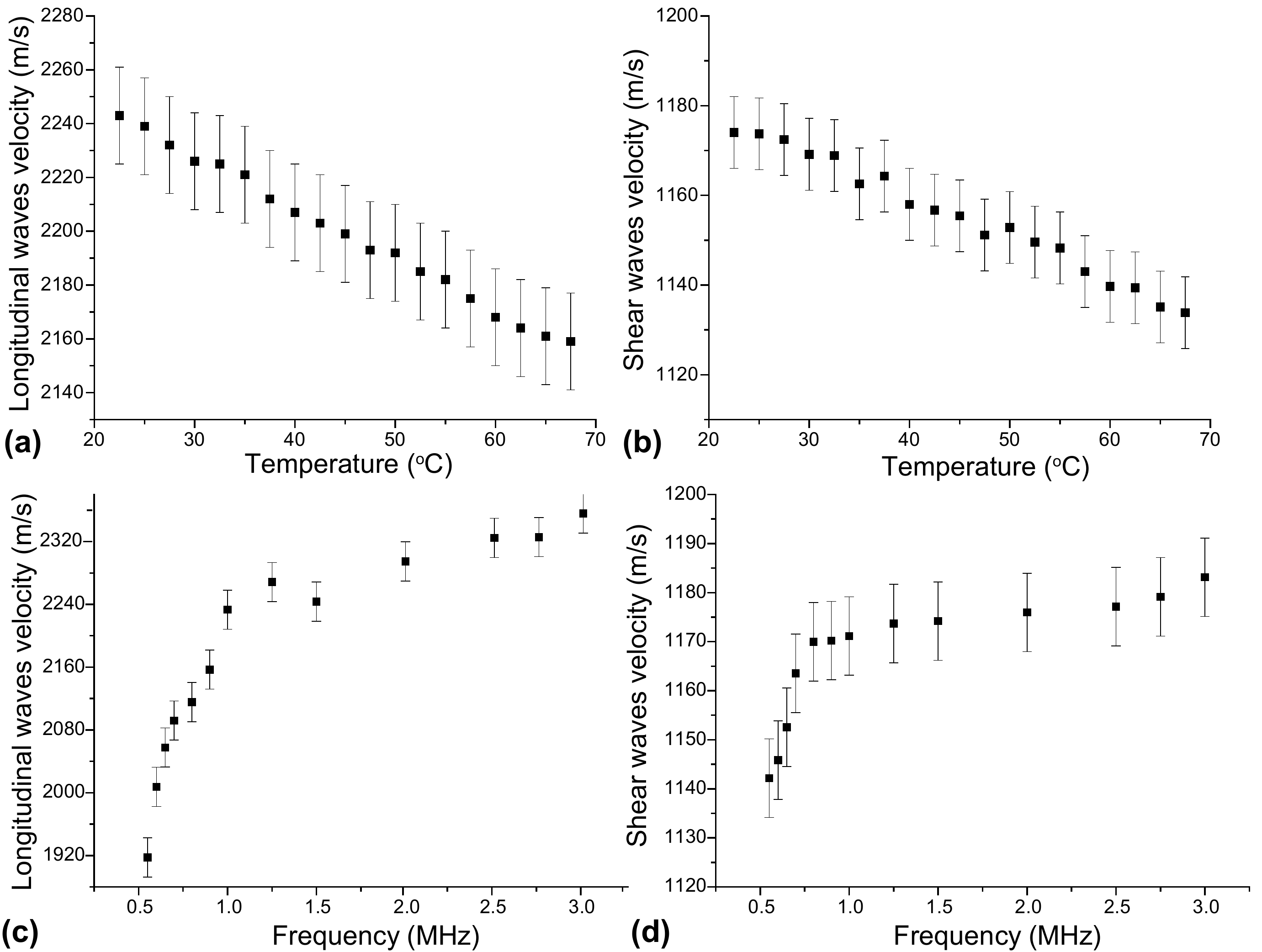}
\caption{Velocities of longitudinal and shear ultrasonic waves as a function of temperature at the frequency of 1.5 MHz (a,b) and as a function of frequency at the temperature of 25 $^\circ$C (c,d).}
\label{velocity}
\end{figure}

Figure \ref{velocity} presents the obtained dependencies of velocities of longitudinal and shear ultrasonic waves on temperature at the frequency of 1.5 MHz (Fig. \ref{velocity} (a,b)) and on frequency at the temperature of 25 $^{\circ}$C (Fig. \ref{velocity} (c,d)).
The data on longitudinal ($V_p$) and shear ($V_s$) wave velocities measured at different temperatures $V_i(T)$ were used for estimating the effective elastic moduli $M_j(0,T) = \rho V_i^2(T)$ and further for calculating the Lam\'e elastic moduli $\lambda$ and $\mu$ using Eqs.~(\ref{eq:Mx}) and (\ref{eq:My}). 
The data obtained are summarized in Table \ref{linear}. As can be seen from Fig. \ref{velocity} and Table \ref{linear} both the velocities of longitudinal and shear ultrasonic waves and linear elastic moduli demonstrated a monotonous decrease with temperature. This behavior is in good agreement with data previously published elsewhere \cite{Hughes1950,Lamberson1972} and can be explained by rising viscosity of the material at elevating temperatures.

In the given temperature range, the linear elastic moduli decreased linearly with the temperature with slopes $b_\lambda = -5.0 \pm 0.5$ MPa/$^{\circ}$C and $b_\mu = -2.2 \pm 0.5$ MPa/$^{\circ}$C for $\lambda$ and $\mu$, respectively.

\begin{table} [h!]
\caption{Ultrasonic wave velocities at the frequency of 1.5 MHz and linear elastic moduli $\lambda$ and $\mu$ of the polystyrene sample measured at different temperatures} 
\centering
{\begin{tabular}{lccccc} \hline
Temperature & $V_p$ (m/s)  & $V_s$ (m/s)  & $\rho$ (g/cm$^3$)  & $\mu$ (GPa)  & $\lambda$ (GPa)   \\ 
\hline
25 $^\circ$C  & 2240$\pm$18 & 1173$\pm$8 & 1.06 & 1.46$\pm$0.01 & 2.41$\pm$0.01 \\
35 $^\circ$C  & 2220$\pm$18 & 1162$\pm$8 & 1.06 & 1.43$\pm$0.01 & 2.36$\pm$0.01 \\
45 $^\circ$C  & 2200$\pm$18 & 1155$\pm$8 & 1.06 & 1.41$\pm$0.01 & 2.31$\pm$0.01 \\
55 $^\circ$C  & 2182$\pm$18 & 1148$\pm$8 & 1.06 & 1.39$\pm$0.01 & 2.26$\pm$0.01 \\
65 $^\circ$C  & 2160$\pm$18 & 1135$\pm$8 & 1.06 & 1.37$\pm$0.01 & 2.21$\pm$0.01 \\
\hline
\end{tabular}}
\label{linear}
\end{table}

\subsection{Temperature dependence of nonlinear elastic moduli}

The prominent dependence of nonlinear elastic moduli of polystyrene on frequency of ultrasonic waves has been demonstrated recently in our paper  \cite{Belashov-frequency-dependence-nonlinear-2024}, and was explained by viscoelastic properties of the material. The absolute values of nonlinear elastic moduli of polystyrene were shown to rise considerably with decreasing frequency. For instance, the difference by an order of magnitude has been observed in values of $l$ modulus in measurements taken at 3 MHz and 500 kHz. Changes in viscoelastic properties of polystyrene with temperature are supposed to affect the nonlinear elastic moduli as well. These changes were analyzed in this paper at four frequencies of ultrasonic waves of 0.7, 1.0, 1.5 and 3.0 MHz.

\begin{figure}
    \centering
    \includegraphics[scale=0.65]{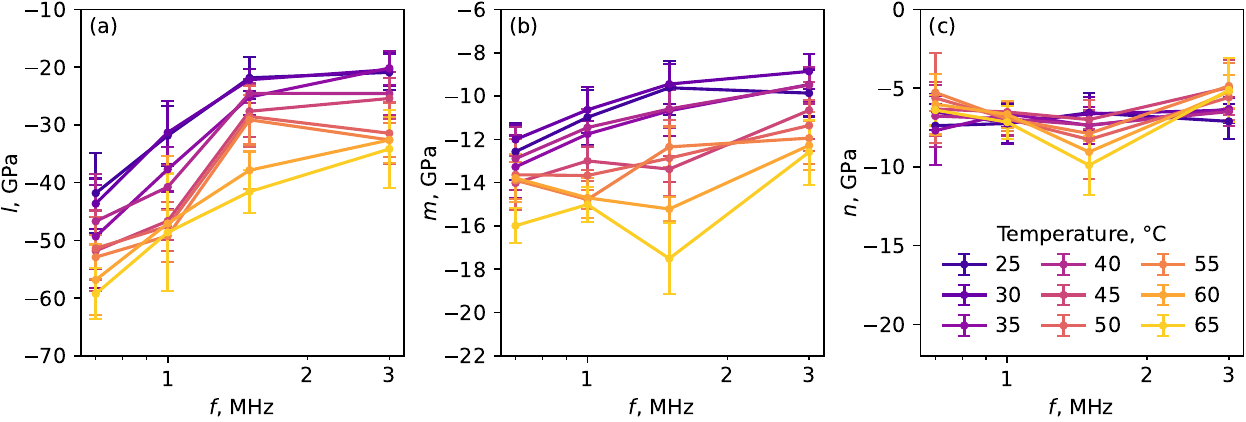}
    \caption{Nonlinear moduli as a function of frequency for different temperatures.}
\label{fig:Nlin_temp}
\end{figure}

The determined above values of velocities of longitudinal and shear ultrasonic waves at different temperatures and frequencies and the obtained values of Lam\'e moduli $\lambda$ and $\mu$ were further used to estimate the nonlinear elastic moduli using the Eqs.~(\ref{eq:l})--(\ref{eq:n}). The results obtained are plotted in Fig.~\ref{fig:Nlin_temp}.

It can be seen that the nonlinear moduli $l$ and $m$ increased in their absolute values with temperature. This temperature dependence is much larger than that for the linear moduli shown in Table~\ref{linear}. For example, the modulus $l$ at frequencies of 1.5--3 MHz doubled at  temperature increasing from 25 to 65 $^\circ$C. At the same time, the modulus $n$ was weakly dependent on temperature and frequency and had a large relative measurement error. Indeed, according to Eq.~(\ref{eq:n}), the modulus $n$ depends on the difference $\alpha_y - \alpha_z$, which is proportional to a small difference in the sensitivities of the two shear wave velocities to the applied pressure.

\begin{figure}
    \centering
    \includegraphics[scale=0.65]{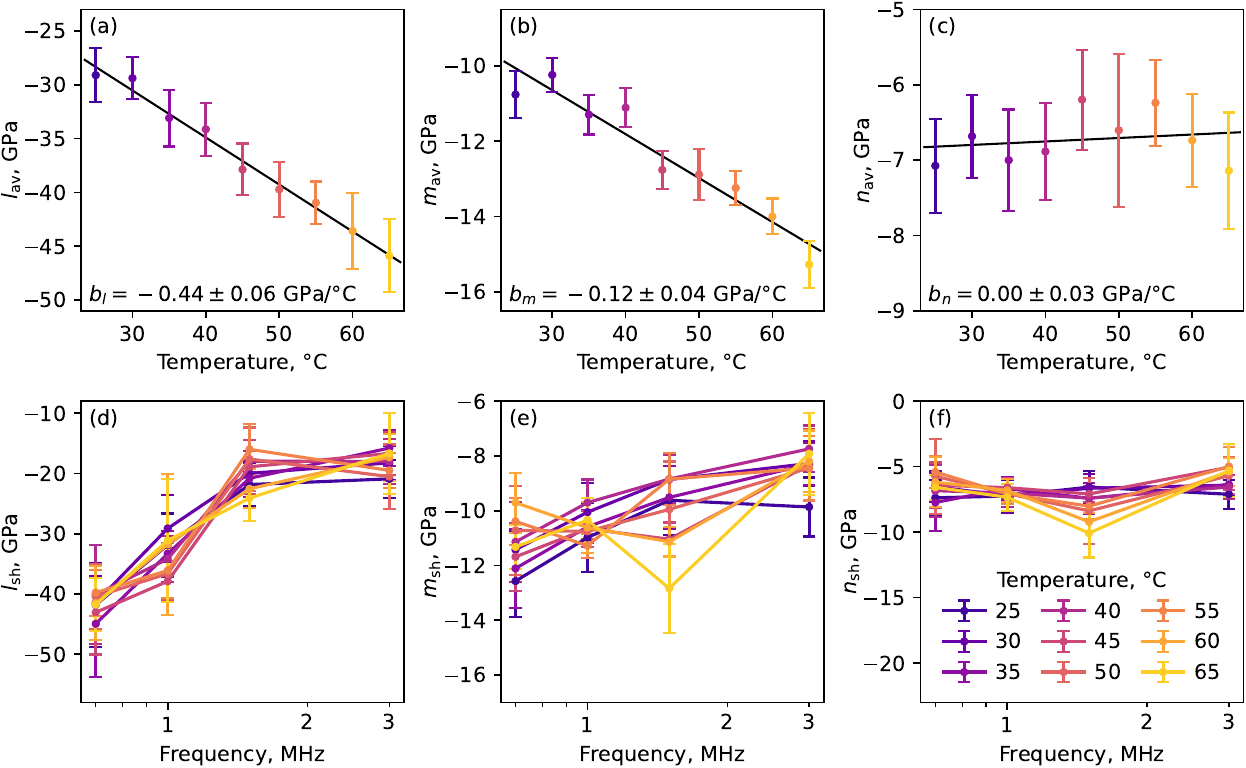}
    \caption{a--c) Averaged nonlinear moduli as a function of temperature. d--f) Shifted nonlinear moduli as a function of frequency for different temperatures.}
\label{fig:Nlin_temp_shifted}
\end{figure}

We are primarily interested in the overall temperature dependence of the nonlinear moduli. For this purpose, we calculated the nonlinear moduli $l_{\rm av}$, $m_{\rm av}$, $n_{\rm av}$ averaged over the measured frequencies for each given temperature.
The resulting dependence of $l_{\rm av}$, $m_{\rm av}$, $n_{\rm av}$ as a function of temperature is shown in Fig.~\ref{fig:Nlin_temp_shifted}(a--c). One can see the linear dependence on the temperature for $l_{\rm av}$, $m_{\rm av}$. For $n_{\rm av}$, there is almost no dependence on  temperature within the measurement tolerance. As a result, we obtained the following slopes $b_l = - 0.44 \pm 0.06$ GPa/$^\circ$C, $b_m = - 0.12 \pm 0.04$ GPa/$^\circ$C, and $b_n = 0.00 \pm 0.03$ GPa/$^\circ$C for the temperature dependence of $l_{\rm av}$, $m_{\rm av}$, and $n_{\rm av}$, respectively. These slopes represent the temperature susceptibility of the nonlinear elastic moduli.

In general, the nonlinear moduli may have different temperature susceptibilities for each frequency. Therefore, we introduce the shifted nonlinear moduli using the previously obtained frequency-independent slopes:
\begin{align}
    l_{\rm sh} &= l - b_l(T - T_0), \\
    m_{\rm sh} &= m - b_m(T - T_0), \\
    n_{\rm sh} &= n - b_n(T - T_0),
\end{align}
where $T_0 = 25$ $^\circ$C is the reference temperature. The obtained values are plotted in Fig.~\ref{fig:Nlin_temp_shifted}(d--f). One can see a good coincidence of the curves, which means that the temperature susceptibilities of the nonlinear moduli are weakly dependent on the frequency in the studied frequency range.



\section{Discussion}

The nonlinear elastic moduli depend on both frequency and temperature. The frequency dependence has been studied in more detail in our previous work~\cite{Belashov-frequency-dependence-nonlinear-2024}. The observed strong dependence of the nonlinear moduli on frequency was explained by a secondary relaxation process in the sub-MHz range, which is sensitive to the static pressure. The molecular origin of this relaxation process remains controversial, but it may be related to the relaxation of some segments of the polymer chain. Since the relaxation processes have different influence on the sound velocity at different frequencies, one obtains a strong frequency dependence of the nonlinear elastic moduli. 

In this paper, we observed a strong dependence of the nonlinear elastic moduli not only on the frequency but also on the temperature. This behavior can be explained by the different influence of  pressure on the relaxation processes at different temperatures. The temperature can shift the relaxation time as well as the mechanical coupling with the pressure.

Qualitatively, the dependence of nonlinear moduli on temperature can be explained by the time-temperature superposition principle. At elevated temperatures, all relaxation processes in the polymer are accelerated. This results in a shift of the frequency dependence towards higher frequencies. Since the absolute values of the nonlinear moduli decrease with frequency, they increase with temperature.  However, the limited frequency range and the measurement errors did not allow us to distinguish the frequency shift from the intrinsic dependence of the nonlinear moduli on temperature. Moreover, the time-temperature superposition principle is reliable only for thermodynamically simple systems, which is rarely the case for systems below the glass transition temperature. Further study is needed to elaborate on this question.

The thermal susceptibility of the nonlinear moduli (except modulus $n$) is two orders of magnitude greater than the temperature susceptibility of the linear moduli. The same order of susceptibility ratio has been found for vitreous silica~\cite{Wang-temperature-dependences-thirdorder-1992} and metal-matrix composites~\cite{Mohrbacher-temperature-dependence-thirdorder-1993}. 


\section{Conclusions}

We have studied the temperature dependence of the nonlinear elastic moduli of polystyrene using the acousto-elastic effect. We have estimated the temperature susceptibility of linear and nonlinear elastic moduli in the temperature range of 25--65 $^\circ$C. 

The temperature susceptibility of nonlinear moduli $l$ and $m$ was two orders of magnitude higher than that of linear moduli $\lambda$ and $\mu$. For the nonlinear modulus $n$, the temperature variation was less than the measurement error and perhaps of the same order as for $\lambda$ and $\mu$. 
The nonlinear moduli $l$ and $m$ were strongly frequency dependent in the range of 0.7--3 MHz for all investigated temperatures. At the same time, the temperature susceptibility of these moduli was weakly dependent on frequency.



\vspace{6pt}

\funding{The financial support from Russian Science Foundation under the grant \# 22-72-10083 is gratefully acknowledged, \url{https://rscf.ru/en/project/22-72-10083/}.}


\dataavailability{Data is available from the authors upon reasonable request.}

\conflictsofinterest{The authors declare no conflicts of interest.} 





\reftitle{References}



\bibliography{sample}

%


\PublishersNote{}

\end{document}